# Robust charge-density wave strengthened by electron correlations in monolayer 1T-TaSe$_2$ and 1T-NbSe$_2$

Yuki Nakata[1], Katsuaki Sugawara[1,2,3], Ashish Chainani[4], Hirofumi Oka[3], Changhua Bao[5], Shaohua Zhou[5], Pei-Yu Chuang[4], Cheng-Maw Cheng[4], Tappei Kawakami[1], Yasuaki Saruta[1], Tomoteru Fukumura[6], Shuyun Zhou[5,7], Takashi Takahashi[1,2,3], and Takafumi Sato[1,2,3]*

[1]Department of Physics, Graduate School of Science, Tohoku University, Sendai 980-8578, Japan.
[2] Center for Spintronics Research Network, Tohoku University, Sendai 980-8577, Japan.
[3]Advanced Institute for Materials Research (WPI-AIMR), Tohoku University, Sendai 980-8577, Japan.
[4]National Synchrotron Radiation Research Center, Hshinchu, 30077, Taiwan ROC.
[5]State Key Laboratory of Low Dimensional Quantum Physics and Department of Physics, Tsinghua University, Beijing 100084, China.
[6]Department of Chemistry, Graduate School of Science, Tohoku University, Sendai 980-8578, Japan.
[7]Frontier Science Center for Quantum Information, Beijing 100084, China
*Corresponding author. E-mail: t-sato@arpes.phys.tohoku.ac.jp

**ABSTRACT**

Combination of low-dimensionality and electron correlation is vital for exotic quantum phenomena such as the Mott-insulating phase and high-temperature superconductivity. Transition-metal dichalcogenide (TMD) 1T-TaS$_2$ has evoked great interest owing to its unique nonmagnetic Mott-insulator nature coupled with a charge-density-wave (CDW). To functionalize such a complex phase, it is essential to enhance the CDW-Mott transition temperature $T_{\text{CDW-Mott}}$, whereas this was difficult for bulk TMDs with $T_{\text{CDW-Mott}}$ < 200 K. Here we report a strong-coupling 2D CDW-Mott phase with a transition temperature onset of ~530 K in monolayer 1T-TaSe$_2$. Furthermore, the electron correlation derived lower Hubbard band survives under external perturbations such as carrier doping and photoexcitation, in contrast to the bulk counterpart. The enhanced Mott-Hubbard and CDW gaps for monolayer TaSe$_2$ compared to NbSe$_2$, originating in the lattice distortion assisted by strengthened correlations and disappearance of interlayer hopping, suggest stabilization of a likely nonmagnetic CDW-Mott insulator phase well above the room temperature. The present result lays the foundation for realizing monolayer CDW-Mott insulator based devices operating at room temperature.

**Introduction**

The interplay among electron correlation, dimensionality, and appearance of various quantum phases is a long-standing issue in condensed-matter physics. The correlated electron system is characterized by strong Coulomb interactions among electrons and the resultant emergence of exotic physical properties, which are absent in the weakly-interacting electron system. The most drastic consequence of electron correlation is the typical Mott-Hubbard transition [1, 2] that converts a half-filled paramagnetic metal (predicted by single-particle theory) into an antiferromagnetic insulator when the on-site Coulomb interaction $U$ exceeds the bandwidth $W$ (*i.e.* effective Coulomb interaction $U/W > 1$). A more unusual phase is the nonmagnetic Mott-insulator and associated exotic quantum phases, as highlighted by the quantum-spin-liquid phase in 1T-TaS$_2$, a triangular lattice of two-dimensional (2D) Mott insulator with a CDW [3, 4]. In comparison, the destruction of antiferromagnetic order in doped copper oxides leads to emergence of high-temperature superconductivity which coexists with charge order [5, 6].

To realize a Mott insulator, it is essential that the magnitude of $U/W$ is above a critical value relevant to the structure and the electronic states of a material [1, 2]. In fact, for an optimally doped copper oxide, the large $U$ ( ~3 eV) [7] estimated for Cu-3$d$ electrons compared with the relatively small bandwidth $W$ ( ~0.4 eV) [8] also satisfies the condition of $U/W \gg 1$. This suggests a direct relation between Mott physics and superconductivity (note that, in cuprates, the role of $U$ is actually played by the charge-transfer gap of 1.4-2.0 eV [9], but even in this case, the effective $U/W$ (3.5-5.0) exceeds the critical value). However, the recent discovery of a Mott-insulator phase and associated superconductivity in tilted bilayer graphene [10, 11] demonstrated that even when $U$ is considerably small (~ 30 meV), the band narrowing ($W$ ~ 20 meV) introduced by the superstructure of moiré pattern can effectively convert a metallic state into a Mott-insulating one. This points to the importance of bandwidth control for materials with small $U$ to trigger the Mott transition.

The layered transition-metal dichalcogenide (TMD) 1T-TaS$_2$ is believed to be a special example of a bandwidth-controlled Mott-transition material [12,13] in the absence of magnetic order. Bulk 1T-TaS$_2$ undergoes a Mott transition accompanying a commensurate charge-density wave (CDW) characterized by the star-of-David cluster (Fig. 1a) with a √13 × √13 periodicity (Fig. 1b), at $T_{\text{CDW-Mott}}$ ~ 200 K. It is noted that twelve Ta atoms located at the corners of a cluster are slightly displaced from the original position towards the central Ta atom (Fig. 1a). 1T-TaS$_2$ satisfies the half-filling condition necessary for realizing a Mott-insulator phase, since 12 electrons at the displaced 12 Ta atoms form the fully occupied 6 bands and the remaining electron at the central Ta atom forms a half-filled metallic band [14, 15]. Although the $U$ of Ta $5d$ electrons is relatively small (~ 0.7 eV) [16], 1T-TaS$_2$ undergoes the Mott-transition when the half-filled band is narrowed to the scale of $U$ due to the band folding associated with the CDW [12, 13] in a similar manner to tilted bilayer graphene. More interestingly, it was shown that while it has a charge gap of ~0.3 eV, it shows gapless quantum spin liquid dynamics and no long range magnetic order down to 70 mK [17]. Recently, the exploration for Mott phases coexisting with CDW was extended to the atomic-layer limit in TMDs as in graphite (graphene), with the possible emergence of exotic quantum phenomena in the pure 2D limit [18-20]. However, the nature of a pure 2D CDW-Mott phase, such as its robustness, possibility for magnetism, and differences if any, compared with the 3D bulk case has been scarcely explored experimentally. In particular, the essential issue regarding the interplay between the Mott phase and dimensionality is yet to be clarified.

In this work, we address all the above key issues by performing a comprehensive angle-resolved photoemission spectroscopy (ARPES) study on epitaxially grown monolayer 1T-TaSe$_2$ and 1T-NbSe$_2$, and demonstrate the robust CDW-Mott phase under external perturbations such as heating and electron doping.

**Results and discussion**

**Characterization of TaSe$_2$.** First, we discuss the electronic structure of monolayer 1T-TaSe$_2$ whose monolayer nature was confirmed by our scanning tunneling microscopy (STM) measurement (Supplementary note 1). Figure 1c displays the 3D ARPES intensity plotted as a function of 2D wave vector ($k_x$ and $k_y$) and binding energy $E_B$ measured at $T$ = 40 K. One can clearly recognize a nearly flat band at $E_B$ ~ 0.3 eV and dispersive holelike bands topped at the Γ point, which are ascribed to the Ta $5d$ and Se $4p$ bands, respectively [18]. The topmost Ta $5d$ band does not cross the Fermi level ($E_F$) and exhibits an insulating gap of ~ 0.3 eV below $E_F$ at the Γ point. This gap is not assigned to a band gap, a substrate-induced gap, or a conventional CDW gap (Supplementary note 2), but to a Mott-Hubbard gap. This Mott gap is associated with the enhancement of $U/W$ caused by the hybridization of backfolded bands and the resultant band narrowing due to the √13×√13 commensurate CDW (Fig. 1b), as in bulk nonmagnetic 1T-TaS$_2$ [12, 13] and a surface layer of bulk 1T-TaSe$_2$ [21, 22] as can be suggested from the overall similarity of experimental band dispersion (Supplementary Fig. S2). The gap size below $E_F$, called here $\Delta_{Mott}$, roughly corresponds to a half of the full Mott-gap size $2\Delta_{Mott}$ because $E_F$ is nearly located at the midpoint between the lower Hubbard band (LHB) and the upper Hubbard band (UHB) as suggested from the comparison of ARPES and tunneling spectroscopy data [20, 23, 24] (Supplementary note 3). As shown in Fig. 1c, a signature of the CDW is clearly seen as an apparent hybridization gap discontinuity [21, 22] in the band dispersion at $k \sim 2/3$ ΓM (red dashed line, see Fig. 1c). It is important to note that we could selectively fabricate a pure 1T-TaSe$_2$ phase (and also 1T-NbSe$_2$ phase, discussed later) with ease by controlling the substrate temperature [18]. This enables observation of a clear hybridization gap discontinuity in our data as compared to a recent study, where admixture from the 1H-TaSe$_2$ phase made it difficult to see the discontinuity [20]. The STM image in Fig. 1d obtained in a spatial region of 8×8 nm$^2$ on a monolayer TaSe$_2$ island signifies a clear periodic modulation associated with the formation of CDW containing the hexagonal lattice of star-of-David clusters. We have confirmed that this lattice has a periodicity of $\sqrt{13} \times \sqrt{13} R 13.9°$ expected for the formation of star-of-David lattice, as well visible as super-spots in the Fourier transform image shown in Fig. 1e, in agreement with the previous STM study of monolayer 1T-TaSe$_2$ [20].

The Ta 4$f$ core-level spectroscopy (Fig. 1f) signifies that the Ta 4$f_{5/2}$ and 4$f_{7/2}$ spin-orbit satellite peaks split into two sub-peaks, as is clearly visible in the energy distribution curve (EDC) at $T = 40$ K. Since the additional splitting of Ta-4$f$ peak is attributed to the different electron density at each Ta site [25, 26] and/or the change in the chemical bonding of Ta atoms due to the formation of the star-of-David clusters, the core-level spectrum is consistent with our STM data which supports the formation of the star-of-David clusters. On elevating temperature, we found that the lower-binding-energy sub-peak of both the Ta4$f_{5/2}$ and 4$f_{7/2}$ components is gradually weakened but the shoulder feature still remains even at $T = 400$ K. This implies that the Mott phase survives much above the room temperature. We will come back to this point later.

**Temperature dependence of the Mott gap.** The formation of CDW is further corroborated by the appearance of a LHB in the ARPES intensity at $T = 300$ K (Fig. 2b), similarly to the case at $T = 40$ K (Fig. 2a), because the Mott gap cannot be formed without the CDW [12, 13]. Intriguingly, the LHB survives even upto $T = 450$ K (the highest temperature in our experimental set-up, see Fig. 2c), whereas the overall spectral feature becomes less clear. Such spectral feature at $T = 450$ K cannot be explained in terms of the absence of Mott gap and the persistence of CDW gap because of the following reason. In bulk TaSe$_2$ [27], the LHB essentially vanishes at room temperature and a large metallic spectral weight emerges at $E_F$, in contrast to the low temperature (70-220 K) data that displays a peak associated with the LHB. Our ARPES data for monolayer 1T-TaSe$_2$ at room temperature resembles that of bulk TaSe$_2$ at low temperature (Fig. 2d), suggestive of the persistence of a Mott gap at $T = 450$ K.

(Supplementary note 4). The robustness of Mott gap is also seen from the detailed temperature dependence of EDC at the Γ point in Fig. 2d. This is in stark contrast to bulk 1T-TaS$_2$ where a metallic Fermi edge is recovered at $T$ = 300 K. Also, this is distinct from bulk 1T-TaSe$_2$ which shows a clear Fermi-edge cut-off even at $T$ = 30 K (Fig. 2d) and hence, we compared the $T$-dependent behavior of the clear gap observed in monolayer 1T-TaSe$_2$ with bulk 1T-TaS$_2$.

To discuss more quantitatively the gap behavior, we plot the binding energy of the leading-edge midpoint of the EDC, $\Delta_{\text{LEM}}$, as a function of temperature for monolayer 1T-TaSe$_2$, and compare it with that for bulk 1T-TaS$_2$. We expect $\Delta_{\text{LEM}}$ to be directly related with the transport gap in the monolayer instead of the spectroscopic gap $\Delta_{\text{Mott}}$ (for the difference between $\Delta_{\text{LEM}}$ and $\Delta_{\text{Mott}}$, see Supplementary note 5). As one can see from the $\Delta_{\text{LEM}}^2$ vs $T$ plot in Fig. 2e, $\Delta_{\text{LEM}}^2$ for monolayer 1T-TaSe$_2$ shows a nearly linear behavior as a function of $T$ near $T_{\text{CDW-Mott}}$, and exhibits a finite value even at 450 K. This nearly linear behavior is also seen in bulk 1T-TaS$_2$ as shown in Fig. 2e, and was also reported for bulk and monolayer 1T-TiSe$_2$ recently [28]. The $T$ dependence of $\Delta_{\text{LEM}}$ is well reproduced by a semi-phenomenological BCS gap equation based on the mean-field approximation (blue solid curve) expressed as, $\Delta_{\text{LEM}}(T)^2 - \Delta_{\text{LEM}}(T_{\text{CDW-Mott}})^2 \propto \tanh^2(A\sqrt{((T_{\text{CDW-Mott}}/T)-1)})$, where $\Delta_{\text{LEM}}$, $T_{\text{CDW-Mott}}$, and $A$ are the binding energy of the leading-edge midpoint, the CDW-Mott transition temperature, and the proportional constant, respectively [28], which was recently used to characterize bulk and monolayer 1T-TiSe$_2$ (note that the observed temperature dependence of EDC can hardly be explained with the thermal broadening; for details, see Supplementary note 4). From the numerical fittings, the transition temperature was estimated to be $T_{\text{CDW-Mott}}$ ~ 530 K for monolayer 1T-TaSe$_2$, and this is much higher than that obtained for bulk 1T-TaS$_2$ (< 200 K; red circles and curves). The present results suggest that the CDW-Mott-transition temperature $T_{\text{CDW-Mott}}$ of monolayer TaSe$_2$ is very high, being drastically higher than that of bulk TaS$_2$ ($T_{\text{CDW-Mott}}$ ~ 200 K) [12] and a surface layer of bulk TaSe$_2$ ($T_{\text{CDW-Mott}}$ ~ 260 K) [27, 29] (note that a consensus has not been reached for the exact $T_{\text{CDW-Mott}}$ value at the surface of TaSe$_2$; Supplementary note 6 and Fig. S5). In contrast, the increase in $T_{\text{CDW}}$ of 1T-TiSe$_2$ in going from bulk ($T_{\text{CDW}}$ ~200 K) to monolayer ($T_{\text{CDW}}$ ~235 K) was small [28, 30]. If we consider $\Delta_{\text{LEM}}$ to be a measure of half the transport gap (since $E_F$ lies in the middle of the gap, as discussed earlier), $2\Delta_{\text{LEM}}(T = 0)/k_B T_{\text{CDW-Mott}}$ ~12, which is significantly larger than the weak-coupling result of ~3.52. Thus, monolayer 1T-TaSe$_2$ can be classified as a strongly-coupled CDW-Mott phase. It is noted that our time-resolved ARPES experiment suggests that the LHB of monolayer TaSe$_2$ survived after photo excitation even when we adopted the maximum pump fluence above which monolayer samples were damaged (this maximum pump fluence is lower than that in bulk TaS$_2$ [31-36]; for details, see Supplementary note 7 and Fig. S6).

**Carrier doping effect to the Mott gap.** Next, we show the robustness of the LHB against carrier doping. Figures 3a-3c shows the evolution of ARPES intensity as a function of potassium (K) coverage $d_K$ [0 (pristine), 3.2 and 6.4 × 10$^{13}$ atoms/cm$^2$] and corresponding second-derivative intensity plots (d-f) (for details of the $d_K$ estimation, see Methods section). Upon K deposition of $d_K$ = 3.2 × 10$^{13}$ atoms/cm$^2$ which corresponds to ~50 % of the star-of-David density (Fig. 3b), the band structure is shifted downward as a whole with respect to pristine TaSe$_2$ (Fig. 3a) due to the electron doping from K atoms. This suggests that the K deposition dopes electron carriers into a whole monolayer film, as in the case of other monolayer TMD films such as TiSe$_2$ where K deposition causes an overall downward shift of the band structure and disappearance of CDW [37]. On further depositing K atoms (Fig. 3c), the spectral feature becomes significantly broad due to the strong electron scattering by the K-induced disorder, whereas a broad hump originating from the LHB is still seen at $E_B$ ~ 0.6 eV in the EDC in Fig. 3g. The LHB and its systematic downward shift are better visualized in the second-derivative-intensity plots in Fig. 3d-3f. A careful look at the intensity for $d_K$ = 6.4 × 10$^{13}$ atoms/cm$^2$ (Fig. 3f) reveals a bright intensity in the vicinity of $E_F$ which originates from a finite Fermi-edge cut-off, as also seen in the EDC (green curve) in Fig. 3g.

To discuss the spectral evolution upon K dosing in more detail, we have analyzed the spectral weight at $E_F$ relative to that of LHB, $I_{EF}/I_{LHB}$, as a function of K coverage $d_K$. As shown in Fig. 3h, the $I_{EF}/I_{LHB}$ value (red circles) does not exhibit a monotonic behavior as a function of $d_K$, showing a minimum at $d_K$ = 3.2 × 10$^{13}$ atoms/cm$^2$. The non-zero value for $d_K$ = 0 atoms/cm$^2$ may be associated with the tail of LHB extending toward $E_F$, as also seen in the EDC (Fig. 3g). We found that the $I_{EF}/I_{LHB}$ value in monolayer for both $d_K$ = 3.2 and 6.4 × 10$^{13}$ atoms/cm$^2$ is larger than that of the CDW-Mott insulator phase ($T$ = 30 K) in bulk TaS$_2$ (blue square), implying a possible metallic behavior. We have confirmed that such a difference between monolayer and bulk is not associated with the difference in the experimental conditions. This can be seen from Fig. 3h where $I_{EF}/I_{LHB}$ at $T$ = 300 K for monolayer 1T-TaSe$_2$ ($d_K$ = 0) obtained with He lamp (green circle) and synchrotron radiation (red circle) well coincide with each other within our experimental uncertainty.

A simple explanation to account for the observed metallic component may be electron occupation of the UHB. Since the energy position of LHB shifts from 0.28 eV to 0.75 eV with K deposition (Fig. 3g), one would expect the UHB to appear below $E_F$ in the $d_K$ = 6.4 × 10$^{13}$ atoms/cm$^2$ sample since the full Mott-gap is estimated to be 0.5 eV (Supplementary note 3).

However, we found no evidence for the prominent peak from the UHB. This is reasonable because the high Coulomb cost $U$ to populate an electron to the UHB does not guarantee the rigid-band-like electron doping. As an alternative possibility, a metallic K component due to the high density of K atoms (which produces an angle-integrated-type background with a weak Fermi-edge cut-off in EDC) or in-gap states (with suppressed quasiparticle intensity and reduced lifetime) could be conceived. The latter possibility is expected from the Hubbard model for the doped Mott insulator and was observed in spectroscopic studies of cuprates [38].

As shown in Fig. 3h, the $I_{EF}/I_{LHB}$ value in the monolayer sample of $d_K = 6.4 \times 10^{13}$ atoms/cm$^2$ is much smaller than that in the bulk counterpart with the fully melted CDW-Mott state at $T = 300$ K (yellow square). This implies that the framework of LHB itself is still maintained in monolayer even when the system likely becomes metallic upon electron doping, in contrast to the bulk counterpart where even a small amount of electron doping breaks the LHB and leads to the occurrence of superconductivity [39-42]. In particular, doping electrons by substitution of magnetic Fe ions in 1T-Fe$_x$Ta$_{1-x}$S$_2$ was shown to result in a dispersive electron band accompanied by a destruction of the LHB even with 1% Fe substitution [40, 41]. A plausible mechanism of such fragileness in bulk is associated with the Fermi-surface nesting condition which is sensitive to the carrier doping, and has been discussed as a primary cause of incommensurate CDW and resultant Mott phase in bulk 1T-TaSe$_2$ [43, 44]. In this regard, the monolayer data are puzzling and surprising because the LHB still survives even when the nesting condition is modified by the electron doping. In any case, the survival of LHB suggests that the electron correlation is still strong even in the doped monolayer sample.

A recent generalized gradient approximation band structure calculation with on-site Coulomb interaction (GGA+U) for isoelectronic 1T-NbSe$_2$ [45] has reproduced the LHB and Mott gap, consistent with the ARPES data. It is noted though, while the GGA+U study and a very recent DFT (density-function-theory) calculation with GGA on monolayer 1T-TaSe$_2$ [46] suggested a spin-1/2 ferromagnetic Mott-insulator phase, our experimental attempt to detect possible ferromagnetism was not successful [since the detection of ferromagnetism by macroscopic magnetization measurement is difficult for monolayer samples, we carried out a very primitive experiment by just putting a strong Nd magnet (magnetic field ~ 500 mT) on top of a film to detect possible attractive force]. A good agreement of the overall band structure in the Mott-insulator phase between monolayer TaSe$_2$ and bulk nonmagnetic 1T-TaS$_2$ [41] may support the nonmagnetic ground state of monolayer TaSe$_2$, although this point needs to be verified in future, e.g. by x-ray magnetic circular dichroism measurement.

**Comparison between TaSe$_2$ and NbSe$_2$.** Now that the survival of LHB in various conditions is established for monolayer TaSe$_2$, next we explore the CDW-Mott phase in a cousin material, monolayer 1T-NbSe$_2$. One can immediately recognize in the ARPES intensity along the ΓK cut in Fig. 4a that the LHB is well seen at $T$ = 40 K in monolayer NbSe$_2$. The LHB survives even at $T$ = 450 K (Fig. 4b), as is also visible in the EDC at the Γ point in Fig. 4c. A side-by-side comparison of the valence-band ARPES intensity along the ΓM cut between monolayer TaSe$_2$ and NbSe$_2$ in Figs. 4d-4e reveals several common features, such as a nearly flat LHB, dispersive holelike Se-4$p$ bands, and a discontinuity of band dispersion at $k_x$ ~ 2/3 ΓM caused by the hybridization-gap opening due to the CDW. These results demonstrate that the robust Mott-insulator phase coexisting with CDW upon heating is a common characteristic for monolayer TaSe$_2$ and NbSe$_2$.

In the following, we discuss why the LHB in monolayer TaSe$_2$ and NbSe$_2$ is robust unlike in the bulk counterpart. Since one of the key parameters to trigger a CDW-Mott transition is the effective Coulomb interaction $U/W$, it is important to discuss the independent roles of how $U$ and $W$ are independently affected on going from the bulk 3D structure to the monolayer 2D case. The effective on-site Coulomb correlation energy can be described by the equation $U = E_\text{I} - E_\text{A} - E_\text{Pol}$ where, $E_\text{I}$ is the ionization energy, $E_\text{A}$ is the electron affinity, and $E_\text{Pol}$ is the polarization energy which arises from screening due to any electronic perturbation such as removing or adding an electron. This screening causes a strong reduction of $U$ compared to the bare Coulomb interaction $U_\text{bare}$ (= $E_\text{I} - E_\text{A}$). For example, based on a one band Hubbard model, $U$ for the Cu site gets reduced to ~ 4 eV in La$_2$CuO$_4$ compared to $U_\text{bare}$ ~20 eV for Cu atom [47]. For 4$d$ and 5$d$ transition metals, $U$ is expected to be still lower. For 1T-TaS$_2$, 1T-TaSe$_2$ and 1T-NbSe$_2$, typical values of $U$ reported in the literature range from ~0.4/0.7 eV (in the GW approximation/DFT-DMFT approximation [16, 51]) to ~2.0/2.8 eV (in GGA + $U$) [20, 48]). Considering the role of screening in monolayer compared to the bulk case, while the intralayer $E_\text{Pol}$ is expected to show negligible changes in the monolayer case, the interlayer $E_\text{Pol}$ would be suppressed as there are no other layers and the interaction with the substrate is weak, resulting in an effective increase in $U$ compared to the bulk.

Similarly, since there is no out-of-plane or inter-layer hopping in the monolayer i.e. the intrinsic bulk interlayer bandwidth $W_\text{out}$ is absent, the net effective bandwidth $W$ will get reduced. It was suggested from the first-principles band-structure calculations that, although the in-plane bandwidth $W_\text{in}$ becomes significantly small (~ 0.08 eV [50]) due to the band reconstruction associated with CDW, the out-of-plane bandwidth $W_\text{out}$ (~0.54 eV [50]) does not suffer from a strong band-narrowing effect because of the in-plane nature of CDW. In this case, the dominant channel to determine the total $W$ is the interlayer hopping (Fig. 5b) (note that the inter- and intra-layer hopping channels do not contribute in an additive way to the bandwidth, but one can still discuss which plays a dominant role). It is thus inferred that the $U/W$ value in the bulk is largely governed by the interlayer hopping and the bulk is located on the verge of the CDW-Mott transition ($U/W \sim 1.3$ [51]; note that $U$ is $\sim 0.7$ eV in TaS$_2$ [16]). On the other hand, in monolayer, the interlayer hopping is intrinsically absent (Fig. 5b) and the net $W$ is simply associated with the intralayer hopping. Thus, both the increase in $U$ and decrease in $W$ are expected to positively work together to efficiently increase $U/W$, leading to the robust CDW-Mott phase far above bulk $T_{Mott}$. We remark here that it is difficult to experimentally determine $W$ by simply tracing the continuously dispersing bands in the experiment, because such bands are composed of multiple subbands reconstructed by the CDW and the intensity of bands is often suppressed in the region away from the original unfolded band [33]. The Mott transition is associated only with a half-filled subband crossing $E_F$ which has a narrow in-plane bandwidth $W_{in}$ [33], although previous ARPES studies above $T_{Mott}$ on bulk TaS$_2$ (e.g. [21, 26, 27, 41]) were unable to resolve this band, probably because of the smearing of fine band structure by e.g. thermal broadening.

A comparison of characteristic energy scales between monolayer TaSe$_2$ and NbSe$_2$ reveals an intriguing aspect of the CDW-Mott phase in the two systems (Fig. 4d and 4e). As shown in the ARPES-derived band dispersions in Fig. 4f and 4g, a half of the full Mott gap, $\Delta_{Mott}$, in TaSe$_2$ ($0.28 \pm 0.02$ eV) is slightly larger than that in NbSe$_2$ ($0.23 \pm 0.02$ eV). Also, the hybridization gap $\Delta_{CDW}$ in TaSe$_2$ ($0.40 \pm 0.03$ eV) is larger than that in NbSe$_2$ ($0.32 \pm 0.03$ eV). According to the general trend of $U$ in $3d$-$4d$-$5d$ electron systems, $U$ for the Nb-$4d$ orbital is expected to be larger than that for the Ta-$5d$ orbital. Band-structure calculations suggested that the in-plane bandwidth of Nb $4d$ band in the normal state of monolayer NbSe$_2$ ($\sim 2.2$ eV) [48] is smaller than that in monolayer TaSe$_2$ ($\sim 2.7$ eV) [52]. It is expected from a simple band-folding picture that the bandwidth in the CDW phase $W_{in}$ is also smaller in NbSe$_2$. Although this argument suggests a larger $U/W$ and a resultant more stable CDW-Mott phase in NbSe$_2$ than in TaSe$_2$, the observed smaller $\Delta_{Mott}$ (0.23 eV) in NbSe$_2$ (this also implies lower $T_{Mott}$) apparently contradicts with the above simple argument. This discrepancy may be reconciled by taking into account the observed larger $\Delta_{CDW}$ (0.40 eV) in TaSe$_2$ which suggests a smaller intralayer hopping and as a result a larger $U/W$ compared to NbSe$_2$ (note that while the bandwidth of LHB is estimated to be 0.19 and 0.21 eV for TaSe$_2$ and NbSe$_2$, respectively, these values cannot be regarded as $W$ and one needs to know the original bare $W$ without any influence from the Mott gap). This suggests that the lattice displacement in the star-of-David cluster is stronger in monolayer TaSe$_2$ (Fig. 5c; left), which is also inferred from the stronger metallic-bonding character of Ta than that of Nb. The stronger periodic lattice distortion due to the CDW in TaSe$_2$ is also supported by the observation of more pronounced hybridization gap discontinuity around $k \sim 2/3$ ΓM in TaSe$_2$, as seen from the stronger intensity suppression at $E_B \sim 0.6$ eV in Fig. 4d than that in Fig. 4e (note that some other experiments must be carried out to firmly establish the stronger CDW in TaSe$_2$). It is also noted that $\Delta_{CDW}$ of monolayer TaSe$_2$ (0.40 eV) is slightly larger than that of bulk TaSe$_2$ (0.37 eV) [22]; this suggests a stronger lattice distortion in monolayer systems, consistent with GGA+U calculations discussed above [45]. All these arguments suggest that the robust CDW-Mott insulator phase of monolayer TaSe$_2$ and NbSe$_2$ (Fig. 5a) are caused by the disappearance of interlayer hopping assisted by a strong lattice distortion. In other words, the robust CDW in monolayer TaSe$_2$ and NbSe$_2$ is derived from a combination of electron correlations, a strong lattice distortion, and the absence of interlayer hopping. It is emphasized that such properties are all linked to the electron hopping (or electron kinetic energy) of the system (Fig. 5b and 5c), and the controllability of the Mottness (*i.e.* strength of the Mott phase) lies on how to effectively manipulate both inter- and intra-layer hopping (Fig. 5a-5d).

The present study has established an effective means to stabilize the CDW-Mott phase in terms of band engineering. Also, the discovery of the robust CDW-Mott phase far above the room temperature is useful for developing practical CDW-Mott insulator-based ultrathin nanoelectronic devices. It would be very interesting to explore the superconductivity in a metallic state near the Mott phase.

**METHODS**

Sample preparation

Monolayer 1T-TaSe$_2$ and 1T-NbSe$_2$ films were grown on bilayer graphene / 6*H*-SiC by using molecular-beam-epitaxy (MBE) method in an ultrahigh vacuum (UHV) of $3\times10^{-10}$ Torr. As for the monolayer NbSe$_2$ film, we have adopted the same procedure to grow monolayer 1*T*-NbSe$_2$ established in our previous studies where the 1T structure, $\sqrt{13}\times\sqrt{13}$ lattice reconstruction and its monolayer nature were already clarified [19, 55]. To fabricate a monolayer TaSe$_2$ film, we have also followed the fabrication method established by ourselves [18]. Specifically, bilayer graphene was prepared by annealing an *n*-type Si-rich 6*H*-SiC(0001) single-crystal wafer, with resistive heating at 1100℃ in ultrahigh vacuum better than $1\times10^{-10}$ Torr for 30 min. A monolayer TaSe$_2$ (NbSe$_2$) film was grown by evaporating Ta (Nb) on the bilayer graphene substrate kept at 560℃ (580°C) under a Se atmosphere [18, 19]. The as-grown film was subsequently annealed at 400°C for 30 min. The growth process was monitored by reflection high-energy reflection diffraction (RHEED). The film thickness was estimated by a quartz-oscillator thickness meter, scanning tunneling microscopy (STM), and atomic force microscopy (AFM). Based on our experience of fabricating various monolayer TMD films such as NbSe$_2$, VTe$_2$, VSe$_2$, and TiSe$_2$ [18, 19, 53, 54], a monolayer film is formed immediately after the disappearance of buffer-layer-originated $6\sqrt{3} \times 6\sqrt{3}$ RHEED pattern upon co-evaporation of transition-metal and chalcogen atoms. We have judged the thickness of 1T-TaSe$_2$ and NbSe$_2$ films by monitoring this disappearance in the RHEED pattern. After the fabrication by the MBE method, the films were transferred to the ARPES-measurement chamber without breaking the vacuum. We have calibrated the deposition rate of K atoms by calculating the volume of π-band Fermi surface at the K point in bilayer graphene on SiC with keeping the same evaporation rate as that in the case of monolayer TaSe$_2$, and it is estimated to be $1.6 \times 10^{13}$ atoms/cm$^2$/min. We have deposited K atoms for 2 and 4 minutes that corresponds to the K coverage $d_K$ of 3.2 and $6.4 \times 10^{13}$ atoms/cm$^2$, i.e. ~50 and ~100 % of the Star-of-David density, respectively. Thus, the amount of K dosing with respect to the Star-of-David density is sufficient to achieve a sizable electron doping into monolayer 1$T$-TaSe$_2$.

ARPES and STM measurements

ARPES measurements were carried out using an MBS-A1 electron-energy analyzer with a high-flux helium discharge lamp and a toroidal grating monochromator at Tohoku University and an Omicron-Scienta R4000 electron-energy analyzer with synchrotron radiation at Taiwan Light Source (TLS), National Synchrotron Radiation Research Center (NSRRC). The energy and angular resolutions were set to be 12.5-40 meV and 0.2°, respectively. Core-level photoemission spectroscopy measurement was performed at BL28A with micro-focused beam spot in Photon Factory. Time-resolved ARPES measurements were carried out at Tsinghua University using an Omicron-Scienta DA30-L-8000 electron-energy analyzer and a Ti: sapphire oscillator which produces femtosecond pulses from 700 to 980 nm at 76 MHz repetition rate with pulse duration of ~ 130 fs. The time resolution was 480 fs when the probe photon energy was set to 6.2 eV [56]. The infrared laser was frequency-quadrupled using BBO and KBBF crystals to produce an ultraviolet probe light from 177.5 to 230 nm. The beam sizes of pump and probe beam were set to ~45 μm and ~15 μm (full width at half maximum), respectively. The wavelengths of pump and probe beam were set to 800 and 200 nm, respectively. The repetition rate was set to 1000 kHz using a pulse picker. The energy and angular resolutions were set to be 8 meV and 0.1°, respectively. The Fermi level ($E_F$) of sample was calibrated with a gold film deposited onto the substrate. To avoid contamination of the sample surface in *ex-situ* ARPES measurements, we covered the film with amorphous Se immediately after the epitaxy, transferred it to a separate ARPES chamber, and then de-capped the amorphous Se film by heating under UHV.

STM measurements were carried out using a custom-made ultrahigh vacuum STM system [57]. Se capping layers for surface protection of TaSe$_2$ films were removed in the STM chamber by Ar$^+$ ion sputtering for 30 min and annealing at 200°C for 60 min. STM measurements were performed with PtIr tips at 4.8 K under UHV below $2.0 \times 10^{-10}$ Torr. All STM images were obtained in constant current mode.

**ACKNOWLEDGMENTS**

We thank Takumi Sato, T. Taguchi, and C.-W. Chuang for their assistance in the ARPES measurements. We also thank NSRRC-TLS for access to beamline BL21B1. This work was supported by JST-CREST (no. JPMJCR18T1), JST-PRESTO (no. JPMJPR20A8), Grant-in-Aid for Scientific Research on Innovative Areas "Topological Materials Science" (JSPS

KAKENHI Grant numbers JP15H05853, and JP15K21717), Grant-in-Aid for Scientific Research (JSPS KAKENHI Grant numbers JP21H04435, JP17H01139), National Natural Science Foundation of China (11725418, 11427903), Ministry of Science and Technology of China (2016YFA0301004, 2015CB921001), Beijing Advanced Innovation Center for Future Chip (ICFC), Tsinghua University Initiative Scientific Research Program, Tohoku-Tsinghua Collaborative Research Fund, Grant for Basic Science Research Projects from the Sumitomo Foundation, Research Foundation of the Electrotechnology of Chubu, Ministry of Science and Technology of the Republic of China, Taiwan under contract no. MOST 108-2112-M-213-001-MY3 and World Premier International Research Center, Advanced Institute for Materials Research. Y. N. and T. K. acknowledges support from GP-Spin at Tohoku University. A.C. and C.M.C. thank the Ministry of Science and Technology (MOST) of Taiwan, Republic of China, for financially supporting this research under Contract No. MOST 109-2911-I-213-501.

**AUTHOR CONTRIBUTIONS**

The work was planned and proceeded by discussion among Y.N, K.S., and T.S. Y.N. and K.S. fabricated ultrathin films. Y.N., K.S., A.C., C.B., S.Z., P.C., C.C., T.K., Y.S., and S.Z. performed the ARPES measurements. H.O. and T.F. performed the STM measurements. Y.N., A.C., T.T., and T.S. finalized the manuscript with inputs from all the authors.

**ADDITIONAL INFORMATION**

**Correspondence** and requests for materials should be addressed to T. S. (e-mail: t-sato@arpes.phys.tohoku.ac.jp)

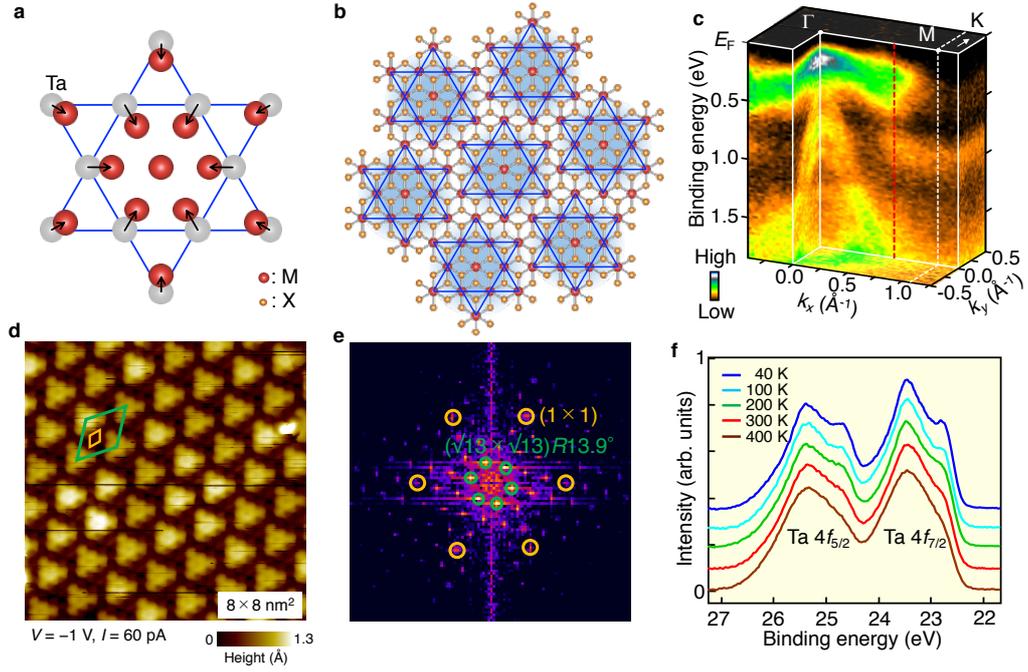

**Fig. 1. Schematics of star-of-David clusters and core-level photoemission spectrum of monolayer 1T-TaSe$_2$.** **a**, Schematics of the displacement of Ta atoms in the star-of-David cluster. M and X represent transition-metal and chalcogen atoms, respectively. **b**, Schematics of crystal structure for monolayer 1T-TaSe$_2$ and star-of-David clusters with the $\sqrt{13}\times\sqrt{13}$ periodicity. **c**, 3D ARPES-intensity plot as a function of 2D wave vector ($k_x$ and $k_y$) and $E_B$ for monolayer 1T-TaSe$_2$ measured at $T$ = 40 K with the He-Iα line ($h\nu$ = 21.218 eV). Hybridization gap ($k_x$ ~2/3 ΓM) is indicated by red dashed line. **d**, STM image in a surface area of 8×8 nm$^2$ for monolayer 1T-TaSe$_2$ on bilayer graphene measured at $T$ = 4.8 K. **e**, Fourier transform image of **d**. **f**, Temperature-dependence of EDC around the Ta-4$f$ core level measured with $h\nu$ = 260 eV for monolayer 1T-TaSe$_2$.

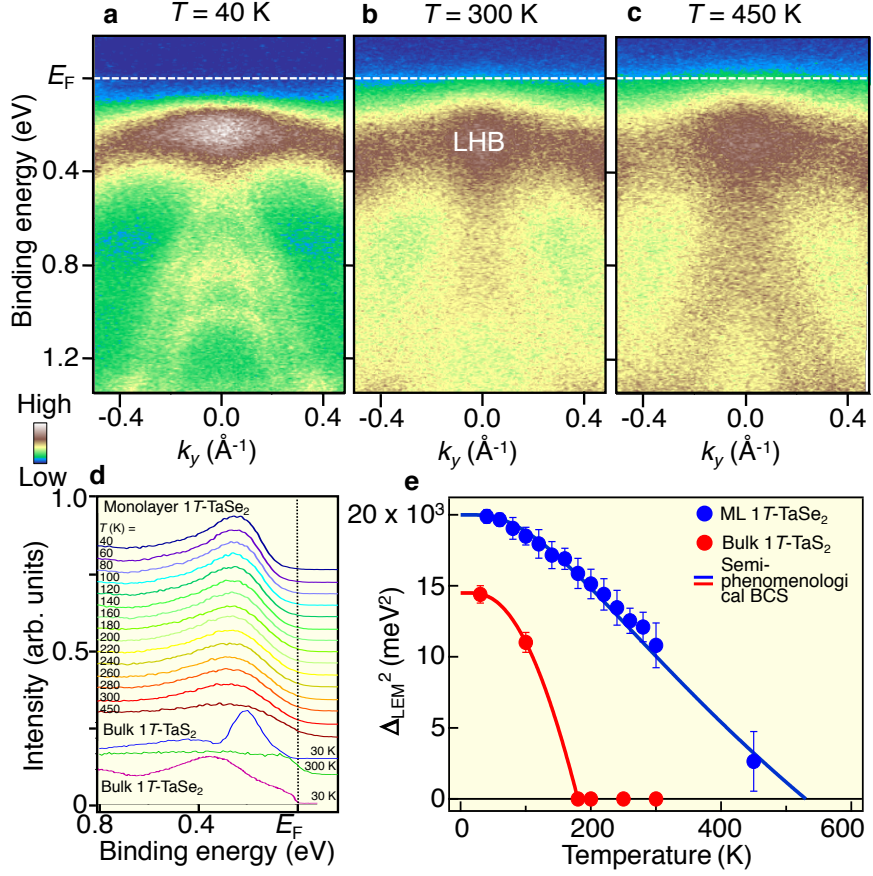

**Fig. 2. CDW-Mott phase of monolayer 1T-TaSe$_2$ robust against temperature variation. a-c**, Near-$E_F$ ARPES intensity along the ΓK cut measured at $T$ = 40, 300, and 450 K, respectively. **d**, Temperature dependence of EDC at the Γ point. EDC for bulk 1T-TaS$_2$ ($T$ = 30 and 300 K) and bulk 1T-TaSe$_2$ ($T$ = 30 K) are also shown as a reference. **e**, squared leading-edge midpoint Δ$_{LEM}$ at the Γ point plotted against $T$ for monolayer 1T-TaSe$_2$ (blue circles), together with the numerical fitting results with the semi-phenomenological (blue solid curve) BCS gap functions. Δ$_{LEM}$ and fitting results are also plotted for bulk 1T-TaS$_2$ (red). We have obtained {$T_{CDW-Mott}$, $A$} = {530 K, 1.01} and {180 K, 1.50} for monolayer 1T-TaSe$_2$ and bulk 1T-TaS$_2$, respectively. Error bars in **e** reflect the uncertainties originating from the energy resolution and the standard deviation in the peak positions of EDCs.

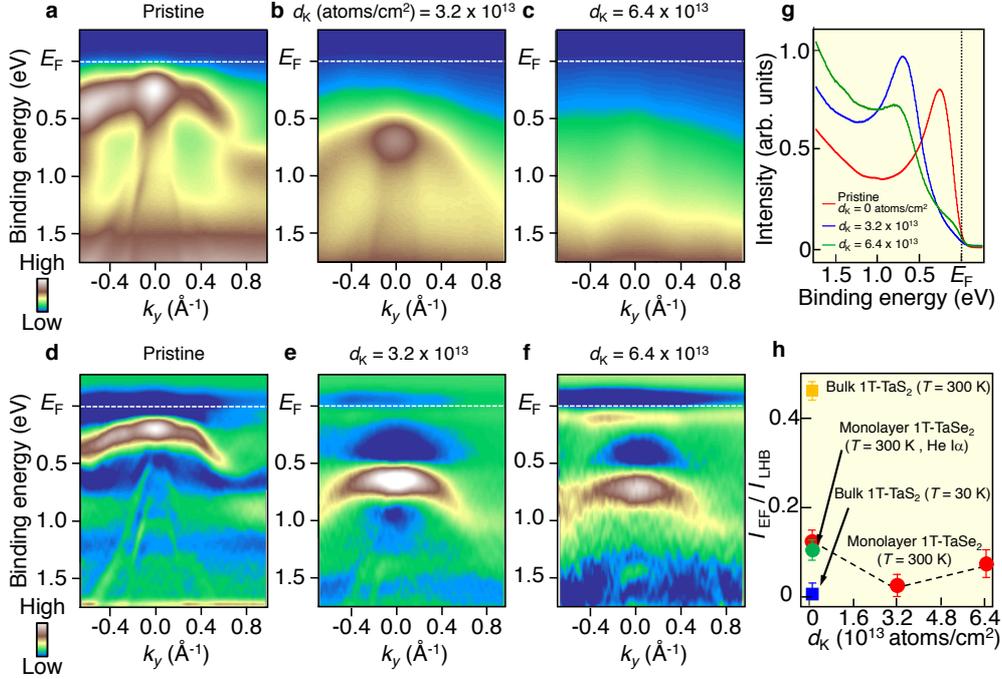

**Fig. 3. CDW-Mott insulator phase in monolayer 1T-TaSe$_2$ robust against electron doping. a-c**, K-deposition dependence of ARPES intensity along the ΓK cut for monolayer 1T-TaSe$_2$ [potassium coverage $d_K$ = 0 (pristine), 3.2 × 10$^{13}$, and 6.4 × 10$^{13}$ atoms/cm$^2$, respectively], measured at $T$ = 300 K with $h\nu$ = 51 eV. **d-f**, Same as **a-c**, but obtained by taking the second derivative of EDCs. **g**, EDCs at the Γ point for each $d_K$. **h**, Plots of intensity at $E_F$ with respect to that at LHB, $I_{EF}/I_{LHB}$, as a function of $d_K$, estimated from the EDCs in **g**. The values for bulk TaS$_2$ measured at $T$ = 30 and 300 K are also plotted. Error bars reflect the uncertainties originating from the energy resolution and statistics of data. Asymmetry in the intensity profile in **a** and **b** is associated with the inequivalent photoelectron matrix-element effect between positive and negative $k_y$'s (Supplementary note 8 and Fig. S7).

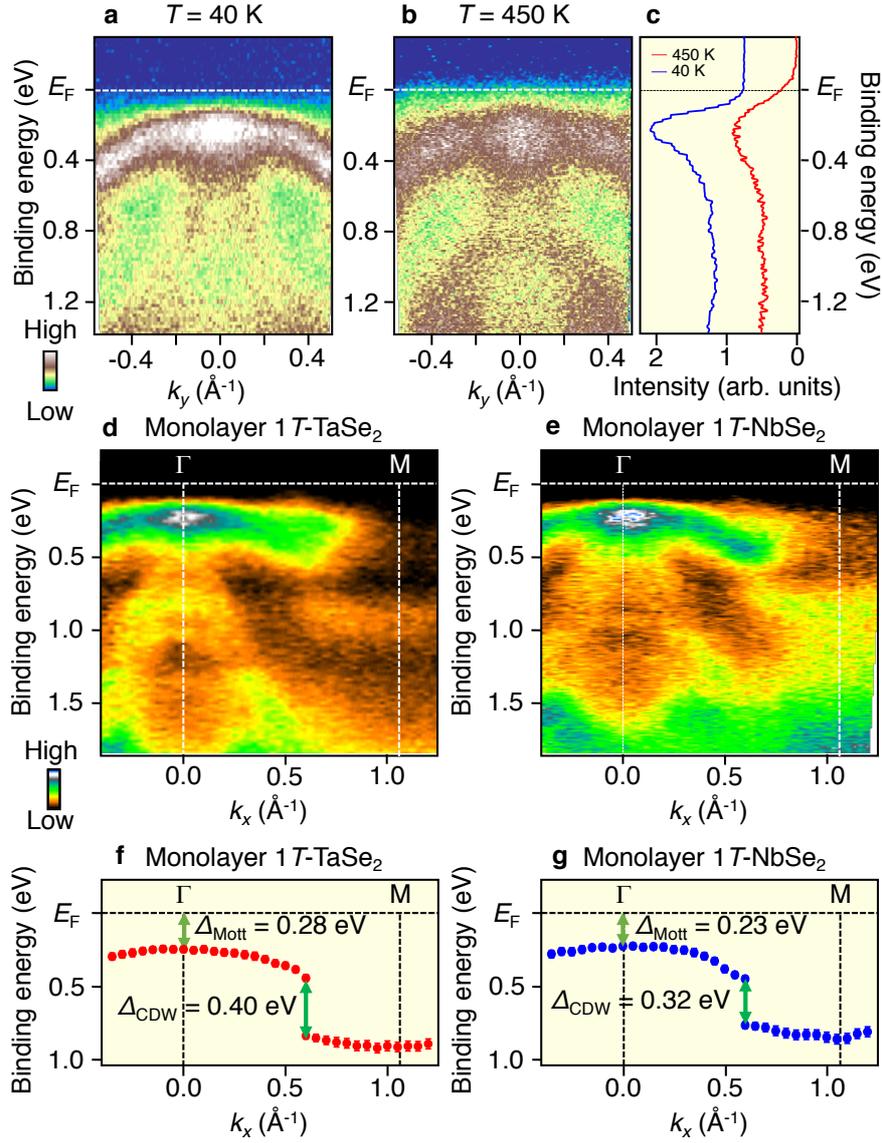

**Fig. 4. Comparison of the CDW-Mott phase between monolayer NbSe$_2$ and TaSe$_2$. a, b** ARPES intensity of monolayer 1T-NbSe$_2$ at $T$ = 40, and 450 K, respectively, measured along the ΓK cut. **c**, EDCs at the Γ point at $T$ = 40 and 450 K. **d, e** Valence-band ARPES intensity along the ΓM cut for monolayer NbSe$_2$ and TaSe$_2$, respectively. **f, g** Experimental band dispersion extracted by tracing the peak position in EDCs for monolayer TaSe$_2$ and NbSe$_2$, respectively, highlighting the quantitative difference in the magnitude of the Mott-Hubbard gap ($\Delta_{Mott}$) and the hybridization gap associated with CDW ($\Delta_{CDW}$). Error bars in **e** reflect the uncertainties originating from the energy resolution and the standard deviation in the peak positions of EDCs.

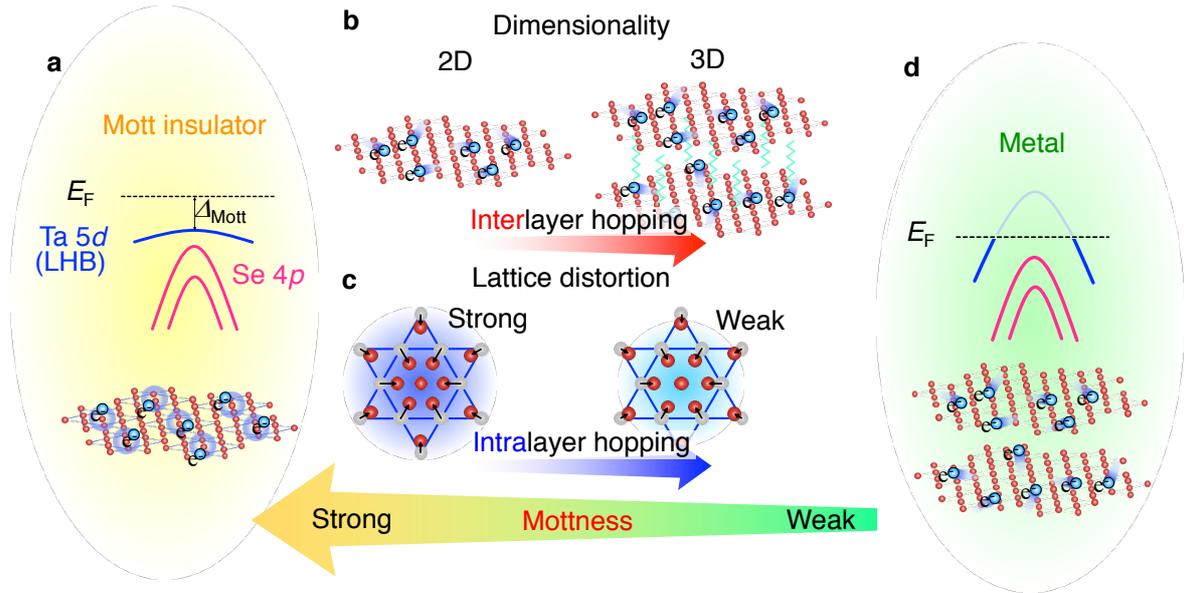

**Fig. 5. Realization of 2D CDW-Mott insulator phase assisted by strong lattice distortion. a**, Schematics of (top) the band dispersion and (bottom) the star-of-David clusters in the robust CDW-Mott phase in monolayer TaSe$_2$. **b**, Illustration to compare the interlayer hopping in the crystal in 2D and 3D systems. **c**, Illustration of strong vs weak lattice distortion in the star-of-David cluster and its relationship with the intralayer hopping in the CDW phase. **d**, Schematics of (top) the normal-state band dispersion and (bottom) the bulk crystal structure without CDW formation.

# SUPPLEMENTARY INFORMATION for

## "Robust charge-density wave strengthened by electron correlations in monolayer 1T-TaSe$_2$ and 1T-NbSe$_2$" by Yuki Nakata *et al.*

**Supplementary note 1: STM characterization of monolayer 1T-TaSe$_2$**

We have performed scanning tunneling microscopy (STM) measurements on the monolayer 1T-TaSe$_2$ film grown on bilayer graphene substrate. As seen from the obtained STM image in a 100×100 nm$^2$ spatial region for the TaSe$_2$ film at $T$ = 4. 8 K in Fig. S1a, a few triangular TaSe$_2$ islands (yellow region) are recognized on top of bilayer graphene substrate (dark region). We have also confirmed from the height profile along a cut across a step of TaSe$_2$ island in Fig. S1b (obtained along red arrow in Fig. S1a) that the step height is ~0.94 nm, which is in between monolayer (0.63 Å) and bilayer (1.26 Å) heights in bulk TaSe$_2$ [1], supporting the monolayer nature of our film.

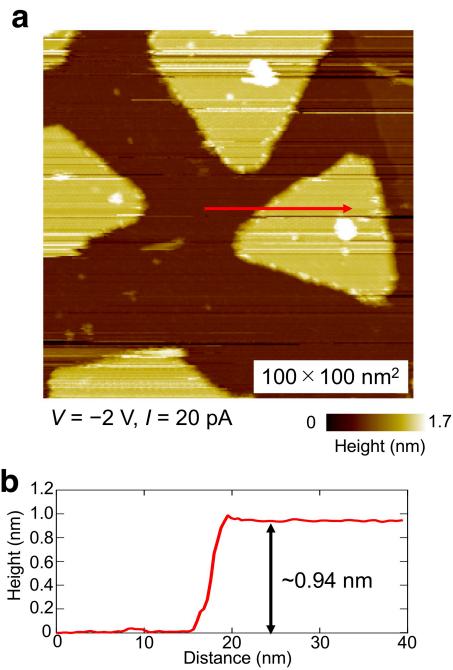

**Supplementary Figure 1: STM characterization of monolayer 1T-TaSe$_2$. a**, STM image in a surface area of 100×100 nm$^2$ for monolayer 1T-TaSe$_2$ on bilayer graphene measured at $T$ = 4. 8 K. **b**, Height profile along a cut across a step of TaSe$_2$ island shown by red arrow in (a).

**Supplementary note 2: Origin of insulating gap in monolayer 1T-TaSe$_2$**

We discuss the origin of an insulating gap (Figs. 1-2) observed in the ARPES experiment of monolayer 1T-TaSe$_2$. Assuming that the observed gap is not a Mott-Hubbard gap, it could arise from any of the following four possibilities; (i) the Anderson gap, (ii) the band gap, (iii) the substrate-induced gap, and (iv) the conventional Fermi-surface-nesting-driven charge-density-wave (CDW) gap. The possibility (i) is ruled out because 1T-TaSe$_2$ is not a strongly disordered system and the experimental density of states (see e.g. Fig. 2d) does not show a power-law-like behavior expected from the disorder-induced Coulomb gap. The possibility (ii) is unlikely because the DFT (density functional theory) calculation of monolayer 1T-TaSe$_2$ neglecting $U$ [2] was unable to reproduce a finite energy gap at the Γ point even when the energy position of the Fermi level ($E_F$) is intentionally moved up or down in the calculation to account for the possible self-doping of carriers to the film and/or the charge transfer from the substrate. The possibility (iii) is also unlikely because of the following reasons. If the gap opens due to the interaction with the substrate, the LHB of TaSe$_2$/NbSe$_2$ is expected to be hybridized with the graphene bands through a direct band overlap. However, this hybridization would not occur because the graphene band is 4 eV away from $E_F$ around the Γ point. Lattice strain by the graphene substrate and resultant change in the band structure are unlikely to be responsible, because the lattice strain is expected to be weak due to the existence of van der Waals gap between TaSe$_2$/NbSe$_2$ and graphene, as suggested from the in-plane lattice constant estimated from the RHEED pattern in monolayer 1T-TaSe$_2$ ($a$ = 3.5 Å [2]) which is similar to that of bulk ($a$ = 3.47 Å [1]). One may think that the gap opening is due to the moiré potential associated with the lattice mismatch between TaSe$_2$/NbSe$_2$ and graphene. But this is also ruled out because the folded subband associated with the moiré potential is not observed. The possibility (iv) is also unlikely because the experiment does not support the emergence of any states within 0.3 eV of $E_F$ that are associated with the CDW-induced backfolded bands predicted in the calculation incorporating the superstructure potential with a √13×√13 periodicity (e.g. [3]). Also, a pseudogap, as observed in the nearly commensurate CCW (NCCDW) phase of bulk 1T-TaS$_2$ [4], is not seen. Based on these arguments, together with the fact that the GGA+U calculation [5] reasonably reproduces the overall experimental band dispersion, in particular, the gap opening at $E_F$, the observed gap is reasonably attributed to the electron-correlation-driven Mott-Hubbard gap (coexisting with CDW). This conclusion is also supported by the comparison of spectral feature between monolayer and bulk. As shown in Fig. S2a-S2c, overall ARPES intensity along the ΓM cut at low temperature appears to be similar between monolayer 1T-TaSe$_2$ and bulk samples of 1T-TaSe$_2$ and 1T-TaS$_2$ in which the Mott-Hubbard nature of the gap is well established. In particular, a nearly flat LHB around the Γ point, as well as the hybridization-gap discontinuity at $k \sim 2/3$ ΓM due to the formation of CDW, are commonly resolved in the ARPES intensity (Fig. S2a-S2c) and the corresponding plots of band dispersion (Fig. S2d-S2f) extracted from the peak position in EDCs.

**Supplementary note 3: Comparison of ARPES and tunneling spectroscopy**

Scanning tunneling microscopy/spectroscopy (STM/STS) experiments on monolayer 1T-TaSe$_2$ [6] reported (i) the local density of states (LDOS) that is asymmetric with respect to $E_F$, and (ii) an anticorrelated behavior in the $dI/dV$-conductance map between the lower and upper Hubbard bands (LHB and UHB). Thus, from the viewpoint of the LDOS and $dI/dV$ maps, the electron-hole symmetry appears to be broken. The authors in ref. 6 have discussed that the anticorrelated behavior distinct from the expectation for an electron-hole symmetric Mott gap is still consistent with the Mott-gap picture and is due to additional effects associated with the tunneling process of electrons. Thus, the anticorrelation behavior in STM may not be against the Mott-insulator nature of the observed gap suggested by ARPES. In addition, the hybridization between the Hubbard bands and other bands may further enhance the particle-hole asymmetric behavior. It is known from previous STM studies that the anticorrelation becomes weak for multilayer (2-3 layer) 1T-TaSe$_2$ and bulk 1T-TaS$_2$ [7, 8]. This does not necessarily indicate that only monolayer sample exceptionally avoids the Mott-insulator behavior. It could rather be explained as: the CDW-Mott insulator behavior realized in monolayer becomes complicated in multilayer and bulk. Such complication may originate from the interlayer hopping, as inferred from the STM data on bulk 1T-TaS$_2$ displaying an intriguing modulation of LDOS around $E_F$ depending strongly on the stacking sequence of David stars along the $c$-axis [8]. In this aspect, comparison of ARPES data between single- and multi-layers would be very interesting. Since the domain size of multilayer islands was found to be rather small, further optimization of epitaxial-growth condition and utilization of a small (submicron) beam spot are required to clarify the band structure of multilayer TaSe$_2$ films.

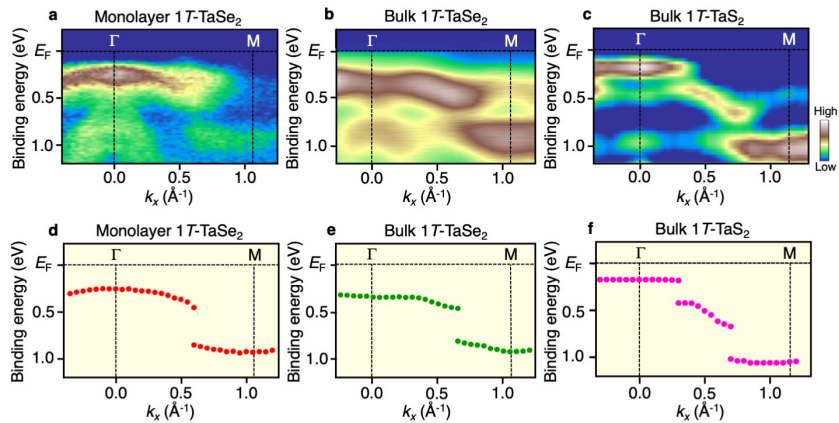

**Supplementary Figure 2: Comparison of band dispersion among monolayer 1T-TaSe$_2$, bulk 1T-TaSe$_2$, and bulk 1T-TaS$_2$. a-c**, ARPES intensity plots along the ΓM cut for monolayer 1T-TaSe$_2$, bulk 1T-TaSe$_2$, and bulk 1T-TaS$_2$, respectively, measured at $T = 30$-$40$ K. **d-f**, Corresponding experimental band dispersion extracted by tracking the peak position in EDCs.

The previous STM study on monolayer 1T-TaSe$_2$ [6] has suggested the gap size of 0.11 eV by defining the width of energy region where the LDOS keeps the zero value, and this value is consistent with our $dI/dV$ curve shown in Fig. S3. We have also estimated the Mott gap value called $\Delta_{Mott}$ from the energy position of LHB relative to $E_F$ at the Γ point in the ARPES spectrum. We found that the energy position of LHB at Γ in the ARPES data nearly corresponds to the maximum of the valence-band peak in the tunneling spectrum in Fig. S3. We have estimated the energy interval between the maxima of valence and conduction bands in the tunneling spectrum to be ~0.5 eV. This value is consistent with the previous STM study [6] and about twice of the $\Delta_{Mott}$ value estimated from the ARPES data (0.56 eV), supporting the non-doped nature of our monolayer 1T-TaSe$_2$ film. Similarly, the non-doped condition is thought to be maintained for our monolayer 1T-NbSe$_2$ sample, as inferred from the previous STS measurement [9] showing that $E_F$ is located at the center of full energy gap (zero density-of-states region).

Since the zero-DOS region is clearly seen around $E_F$ in the STM data for monolayer 1T-TaSe$_2$, it is expected that the electronic transport would also show insulating behavior. However, at the moment, it is difficult to obtain reliable transport data with our monolayer film, because the film is unstable in the atmosphere and therefore not suitable for performing *ex-situ* transport measurements. Even when we cover the sample with a protection layer, the electric current will selectively flow through the metallic bilayer graphene substrate (and the protection layer if it is conductive), and consequently, the insulating behavior of monolayer film cannot be detected. *In-situ* transport measurements would be also difficult because of the same reason. Although the transport measurement is difficult, the zero-DOS behavior in the tunneling spectrum around the zero bias voltage strongly suggests the insulating nature of our monolayer 1T-TaSe$_2$ film.

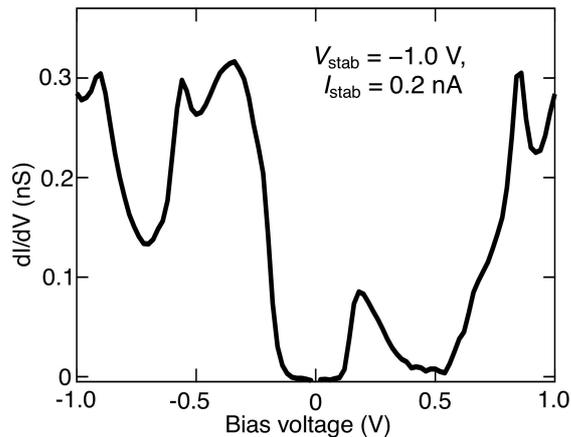

**Supplementary Figure 3: Density of states of monolayer 1T-TaSe$_2$.** Typical $dI/dV$ spectrum on the monolayer TaSe$_2$ island measured at $T = 4.8$ K.

**Supplementary note 4: Temperature evolution of LHB**

It is known from previous studies of bulk 1T-TaS$_2$ that the Mott and CDW states are coupled to each other. Such a coupling manifests as the NCCDW (nearly commensurate CDW) transition below $T_{NCCDW}$ ~ 340 K associated with the formation of local David-star clusters, followed by a Mott transition (metal-insulator transition) at $T_{Mott}$ ~ 200 K accompanying the CCDW (commensurate CCDW) with a √13× √13 periodicity characterized by the fully stacked David-star clusters. Since the Mott transition is likely triggered by the band narrowing that follows the √13× √13 superstructure potential, one can say that the necessary condition (but not the sufficient condition) of the Mott transition is the formation of CDW accompanying the David stars. It is known from the previous literatures [e.g. [4]] that one can distinguish (i) the normal state where both CDW and Mott states are absent, (ii) the Mott-insulator state, and (iii) the NCCDW state, by carefully inspecting the band structure. Specifically, (i) is characterized by the existence of a metallic band forming a large electron pocket centered at the M point as predicted from the first-principles band-structure calculations, (ii) is identified by the appearance of LHB with narrow bandwidth and opening of a full gap on entire Brillouin zone (BZ), and (iii) is confirmed by the absence of LHB and the recovery of nearly normal-state-like band structure with a pseudogap at the Fermi wave vectors ($k_F$'s) on the electron pocket. Spectral distinction of these phases is known to be not so difficult. In particular, the Mott (CCDW) and NCCDW phases can be well distinguished because the transition between the two phases is accompanied by an abrupt change in the overall band structure within the binding-energy range of ~1.2 eV [4].

Based on the above situation in bulk 1T-TaS$_2$, one can think about three possibilities in monolayer 1T-TaSe$_2$, i.e. (i) the CDW appears but the Mott state is not realized, and both CDW and Mott transitions take place (ii) at different temperatures ($T_{NCCDW} > T_{Mott}$) or (iii) at the same temperature ($T_{CDW} = T_{Mott}$). The case (i) is unlikely because we clearly observe a nearly flat band (the LHB) and a full gap on the entire BZ. The case (ii) is characterized by the disappearance of Mott gap and LHB as well as the appearance of CDW-originated pseudogap at $T_{Mott} < T < T_{NCCDW}$. What can be suggested from the present ARPES result is that a nearly flat band seen at 40 K still survives even at 450 K (Fig. 2c), suggestive of the persistence of LHB even at 450 K. This cannot be explained in terms of the absence of Mott gap and the persistence of simple CDW gap at 450 K because the spectral feature continuously evolves in the temperature range of 40 - 450 K, distinct from the abrupt change across $T_{Mott}$ in bulk. Namely, whichever the case (ii) or case (iii) is realized, we can conclude that $T_{Mott}$ for monolayer 1T-TaSe$_2$ is markedly enhanced compared to that for bulk 1T-TaS$_2$. Moreover, since $T_{NCCDW}$ for bulk (~340 K) is much lower than 450 K, it is reasonable to conclude that the CDW transition temperature is also enhanced in monolayer.

We have examined whether or not the observed temperature dependence of the LHB can be explained with the simple thermal broadening effect. As shown in Fig. S4, we have tried to reproduce the EDC at $T$ = 450 K (red curve) by intentionally broadening the EDC at $T$ = 40 K with a gaussian, $\exp(-E^2/2\sigma^2)$, by varying its energy width σ (blue curve). When σ was chosen to match the slope of the leading edge (top curves), the spectral weight around the peak top is significantly overestimated in the simulation. On the other hand, when σ was chosen to match the overall EDC shape relatively well (bottom curves), the simulated curve apparently has a broader leading edge. Also, it is noted that σ values used for the simulation [0.11 (0.18) eV which correspond to $T$ = 1260 (2070) K for the top (bottom) EDC] are too large to be reconciled with a simple thermal effect. These results suggest that the observed temperature dependence of the LHB can hardly be explained with the simple thermal broadening picture, but reflects an intrinsic change in the gap magnitude.

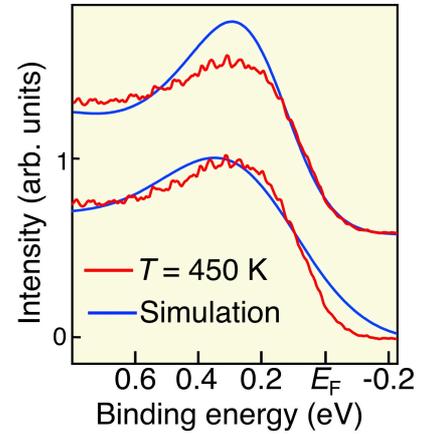

**Supplementary Figure 4: Simulation of thermal broadening effect on LHB.** EDC at the Γ point at $T$ = 450 K (red curve) for 1ML 1T-TaSe$_2$ and simulated EDCs (blue curves) that were generated by broadening the experimental EDC at $T$ = 40 K with a gaussian $\exp(-E^2/2\sigma^2)$ assuming σ = 0.11 eV (top) and 0.18 eV (bottom).

**Supplementary note 5: Difference between the leading-edge gap and spectroscopic gap**

There exist two types of definitions in estimating the gap size from the EDC, i.e. (i) the energy position of the peak relative to $E_F$ and (ii) the energy position of the leading-edge midpoint (LEM) relative to $E_F$. Both definitions are often used to discuss the gap size in other systems such as cuprates and Fe-based superconductors [10, 11]. In the present study, they are named $\Delta_{Mott}$ and $\Delta_{LEM}$, respectively. We call $\Delta_{Mott}$ a spectroscopic gap, because $2\Delta_{Mott}$ spans the energy positions at which the LHB and UHB take the highest (and thereby spectroscopically prominent) density of states (DOS). The DOS for both the LHB and UHB is not so sharp and has a broad tail as can be seen from the ARPES data in Fig. 2d and tunneling spectrum in Fig. S3. Because the thermal excitation starts as soon as the excitation energy exceeds the zero DOS region around $E_F$, the transport measurement is sensitive to this tail. Since the definition of $\Delta_{LEM}$ inherently includes the tail region of the LHB, $\Delta_{LEM}$ is regarded to be sensitive to the transport gap (*i.e.* an activation gap in the transport measurements).

**Supplementary note 6: Comparison of band structure between bulk and monolayer 1T-TaSe$_2$**

We show in Fig. S5 the ARPES intensity plots along the ΓM cut for bulk 1T-TaSe$_2$ measured at $T =$ (a) 30 K and (b) 300 K, compared with (c) that for monolayer 1T-TaSe$_2$ measured at $T =$ 40 K (same as Fig. 4d in the main text). One may see from the comparison of Fig. S5a and S5c that the overall spectral intensity is roughly similar between bulk and monolayer at low temperature, e.g. in the existence of a hybridization-gap discontinuity at $k \sim$ 2/3 ΓM due to the CDW formation. One can also see from the intensity plot at $T =$ 300 K (b) that there exists no apparent band crossing of $E_F$ and the spectral weight at $E_F$ is still suppressed. In this regard, our result does not seem to be fully consistent with the previous study [12] which suggested the surface Mott transition occurring at $T_{Mott} \sim$ 260 K. But this does not necessarily indicate that $T_{Mott}$ at the surface of bulk 1T-TaSe$_2$ is far above room temperature (as high as that of monolayer), because even from the data for bulk at $T =$ 30 K (Fig. S5a) one can clearly recognize a considerable intensity at $E_F$ around the Γ point, in stark contrast to the almost zero intensity at $E_F$ in monolayer (Fig. S5c). Such a crucial difference is highlighted by the comparison of low-temperature EDC at the Γ point in Fig. S5d; the result signifies a clear Fermi-edge cut-off in bulk (blue curve), as opposed to negligible spectral weight at $E_F$ and steep intensity drop for the leading edge of LHB in monolayer (red curve).

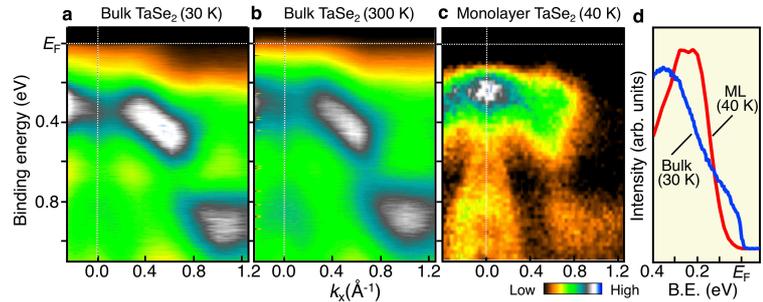

**Supplementary Figure 5: Comparison of band dispersion between bulk and monolayer 1T-TaSe$_2$. a-c**, ARPES intensity plots along the ΓM cut for bulk 1T-TaSe$_2$ measured at $T =$ (**a**) 30 K and (**b**) 300 K, compared with (**c**) that for monolayer 1T-TaSe$_2$ measured at $T =$ 40 K. **d**, EDC at the Γ point for bulk and monolayer 1T-TaSe$_2$, measured at $T =$ 30 and 40 K, respectively.

**Supplementary note 7: Spectral behavior in the photo-excited state**

Figure S6a and S6b show the ARPES intensity around the Γ point measured with a 6.2-eV probe laser before (delay time, $t < 0$) and immediately after ($t = 0$) photo-excitation by a 1.55-eV pump laser, respectively. Taking into account the probe laser energy of $h\nu = 6.2$ eV and assuming that the work function of monolayer 1T-TaSe$_2$ is the same as that of bulk 1T-TaS$_2$ (5.2 eV) [13], one can roughly estimate the upper limit of photoelectron binding energy ($E_B$) to be 1 eV (= 6.2 – 5.2 eV). This $E_B$ well covers the LHB located at ~0.3 eV. Also, a pump laser of $h\nu = 1.55$ eV enables the excitation of electrons to the UHB across the Mott gap of ~0.6 eV. Thus, our time-resolved ARPES measurement fully covers the important energy region involved for the Mott gap. It is noted that photoelectrons are detected at the low kinetic energy region outside the $E_B$ range of 1 eV. Such a spurious photoelectron signal is known to appear because the low kinetic energy cut-off of photoelectron signal is usually broad unless a bias voltage is applied between the sample and analyzer. Also, it is known that photoelectrons with kinetic energy typically lower than a half of pass energy of the analyzer cannot be correctly measured by a standard ARPES apparatus.

As shown in Fig. S6a, at $t < 0$, one can clearly see peaks at $E_B \sim 0.3$, 0.55, and 0.8 eV, which are attributed to the LHB, and two Se-4$p$ bands, respectively, by referring to the experimental result in Fig. 2a measured with the He-Iα line ($h\nu$ = 21.218 eV). While the intensity of LHB is significantly suppressed because of the marked reduction of the Ta-5$d$/Se-4$p$ photoionization cross-section ratio [14], one can still recognize in Fig. S6a a broad peak corresponding to the LHB, as better visualized in the EDC at the Γ point in Fig. S6d (blue curve). After photo-excitation (Fig. S6b), the overall spectral feature becomes broad, but one can still see a weak but finite LHB intensity. This is supported by appearance of a broad hump at $E_B$ ~ 0.3 eV in the EDC in Fig. S6d (red curve), followed by the Se 4$p$ main peak at $E_B$ ~ 0.5 eV.

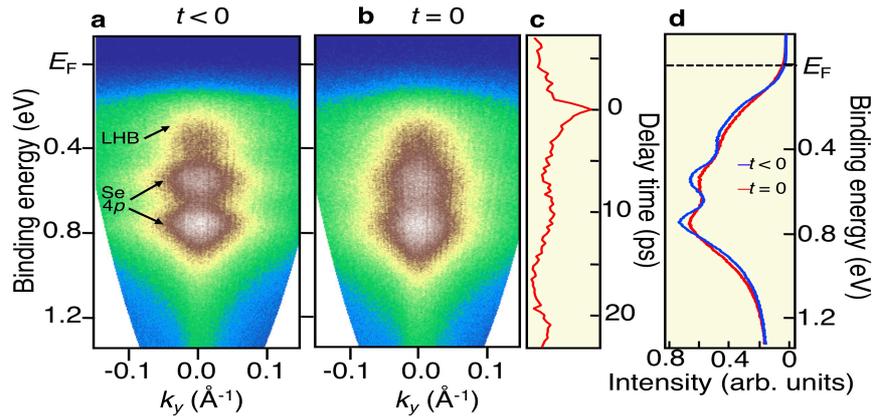

**Supplementary Figure 6: Time-resolved ARPES measurements for monolayer 1T-TaSe$_2$. a** and **b**, ARPES intensity around the Γ point at $T = 80$ K measured with a 6.2-eV probe laser before (delay time, $t < 0$) and immediately after ($t = 0$) photo-excitation by a pump laser of 1.55 eV, respectively. **c**, Delay-time dependence of photoelectron intensity at $E_B = 0.314$ eV (the $E_B$ position of LHB). We set the pump laser power to be as high as possible (0.26 mJ/cm$^2$), above which the monolayer sample showed a rapid and irreversible degradation. **d**,

Considering the high pump fluence to trigger the electronic phase transition in bulk Ta-based TMDs [15-20], we have tried to increase the pump fluence as much as possible, and found that 0.26 mJ/cm$^2$ is an upper limit for reliable measurements, above which an irreversible spectral broadening was observed due to the photo-induced damage of the monolayer sample. Therefore, we carried out the time-resolved ARPES measurement with 0.26 mJ/cm$^2$ pump fluence. On the other hand, there exist some reports that bulk samples are still robust at this pump fluence because the pump fluence higher than 1 mJ/cm$^2$ could be applied without damaging the sample [15-20]. This suggests that the monolayer sample is structurally and/or chemically more fragile against the pump laser irradiation than the bulk counterpart. This may be related to the absence of interlayer coupling in monolayer which may help strengthen the overall stability of sample. Thus, direct comparison of the robustness of CDW-Mott phase against photo-excitations between monolayer and bulk under identical experimental condition is difficult at the moment.

**Supplementary note 8: Geometry of ARPES measurements**

ARPES data shown in Fig. 3 were obtained by using linearly polarized photons from synchrotron radiation while those in Figs. 1 (except for the data in Fig. 1f which was measured with circularly polarized 260-eV photons), 2, and 4 were recorded with the linearly polarized He-Iα photons in laboratory (linear polarization is due to grating). Both data were obtained using the "angular mode" of electron analyzer which simultaneously collects photoelectrons with a finite acceptance angle along the $y$ direction, as shown in Fig. S7. In the case of laboratory-based measurement (Fig. S7a), the light is in $x$-$z$ plane, and the geometry is equivalent for photoelectrons emitted to the positive vs negative $y$ directions. On the other hand, as shown in Fig. S7b, the light is in $y$-$z$ plane in the synchrotron-based measurement, and it is inequivalent. This causes asymmetry in the photoelectron matrix-element effect between positive and negative $k_y$'s, as seen in Fig. 2a of main text. It is noted that the intensity asymmetry does not affect the main conclusion of the present study, because the matrix-element effect does not influence the energy position of bands.

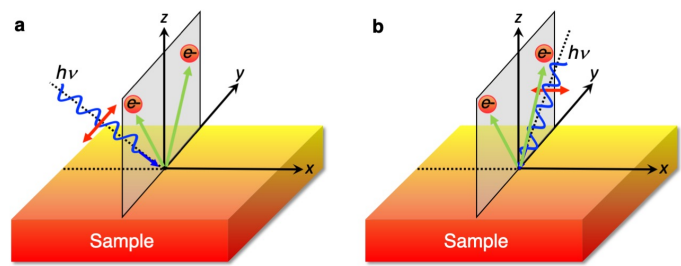

**Supplementary Figure 7: Schematics of experimental geometry for ARPES measurements.** Experimental geometry of incident light, emitted photoelectrons, and sample surface for the **a**, laboratory-based and **b**, synchrotron-based ARPES measurements.

**Supplementary references**